\documentclass[aps,nofootinbib,twocolumn,showpacs]{revtex4-1}
\usepackage{amsmath}
\usepackage{amssymb}
\usepackage{graphicx}

\begin{document}

\title{Proton polarisability contribution to the Lamb shift in muonic hydrogen\\
at fourth order in chiral perturbation theory}
\author{Michael C. Birse and Judith A. McGovern}
\affiliation{Theoretical Physics Division, School of Physics and Astronomy,\\
The University of Manchester, Manchester, M13 9PL, UK\\}

\begin{abstract}
We calculate the amplitude $T_1$ for forward doubly-virtual Compton scattering 
in heavy-baryon chiral perturbation theory, to fourth order in the chiral 
expansion and with the leading contribution of the $\gamma$N$\Delta$
form factor. This provides a model-independent expression for the amplitude in the 
low-momentum region, which is the dominant one for its contribution to the Lamb 
shift. It allows us to significantly reduce the theoretical uncertainty 
in the proton polarisability contributions to the Lamb shift in muonic hydrogen.
We also stress the importance of consistency between the definitions of the Born 
and structure parts of the amplitude. 
Our result leaves no room for any effect large enough to explain the discrepancy
between proton charge radii as determined from muonic and normal hydrogen. 

\end{abstract}

\maketitle

\vskip 10pt

\section{Introduction}

The recent determination of the proton charge radius from the Lamb shift in
muonic hydrogen \cite{pohl10} gives a value that differs by about 5 standard 
deviations from the CODATA value \cite{codata08} and from the results of recent 
electron scattering experiments \cite{a110}. This has generated a flurry of
activity attempting to understand the origin of the discrepancy and whether 
it could be a signal of physics beyond the Standard Model. Recent reviews of 
the situation can be found in Refs.~\cite{jen11a,jen11b,bor12,codata12}.)

However before concluding that new physics is required, it is essential to 
examine carefully any possible conventional explanations. As discussed in 
the reviews, many of the contributions to the Lamb shift are theoretically 
well constrained. One place where some theoretical uncertainty remains is
the contribution of proton structure to two-photon exchange, specifically 
through the polarisability of the proton. 

The energy shift of an $S$-wave hydrogenic state due to two-photon exchange 
can be expressed in terms of the spin-averaged amplitude for forward 
doubly-virtual Compton scattering (V$^2$CS) \cite{bt76}. 
This comprises two tensor structures:
\begin{equation}
\begin{split}
T^{\mu\nu}=&\left(-g^{\mu\nu}+\frac{q^\mu q^\nu}{q^2}\right)
T_1(\omega,Q^2)\cr
&+\frac{1}{M^2}\left(p^\mu-\frac{p\cdot q}{q^2}\,q^\mu\right)
\left(p^\nu-\frac{p\cdot q}{q^2}\,q^\nu\right)T_2(\omega,Q^2),
\end{split}
\label{eq:fv2cs}
\end{equation}
where $p$ and $q$ are the four-momenta of the proton and photon, respectively, 
$M$ is the nucleon mass, $Q^2=-q^2$, and $\omega=p\cdot q/M$.
Dispersion relations \cite{bj73,pach99} can be used to estimate the amplitudes 
$T_{1,2}(\omega,Q^2)$ from the corresponding structure functions measured in 
inelastic electron scattering. These parts of the contribution of proton 
structure are well determined from the available data, with one important
exception: the dispersion relation for $T_1$ does not converge and so it
requires a subtraction. This is normally done at $\omega=0$, introducing a 
dependence on the unmeasured amplitude $T_1(0,Q^2)$.
The slope of this term at $Q^2=0$ is given by a low-energy theorem (LET) in 
terms of the magnetic polarisability of the proton, $\beta$ 
\cite{skk96,dkms97,fs98}. Otherwise its form is unknown and nucleon elastic 
form-factors are often used to model its $Q^2$ dependence 
\cite{pach99,mart06}. The same amplitude is also needed 
in calculations of the electromagnetic self-energies of the nucleons
\cite{zee72,gl82} and it has recently discussed in this context 
\cite{wlcm12}.

This approach to determining the amplitude for forward V$^2$CS has been questioned 
in several recent papers. Miller \textit{et al.}~\cite{mtcr11} 
have suggested that off-shell form-factors of the proton could generate new 
large polarisability contributions to V$^2$CS. They also questioned the validity 
of the LET for the slope. Carlson and Vanderhaeghen \cite{cv11a} estimated the 
$Q^2$ dependence from a two-pion exchange model motivated by chiral perturbation 
theory. They also split the amplitude $T_1(\omega,Q^2)$ into nucleon ``pole" and 
``non-pole" terms, in contrast to Pachucki \cite{pach99}, who separated out the 
full nucleon Born contribution obtained from the Dirac equation with on-shell 
form factors. Finally, Hill and Paz \cite{hp11} used an effective field theory,
nonrelativistic QED, to calculate $T_1(0,Q^2)$, implicitly verifying the LET.
They also questioned the use of on-shell form factors to get the $Q^2$ dependence 
of the non-pole Born terms in $T_1(0,Q^2)$. They did not attempt to estimate the 
polarisability contribution but suggested that the uncertainty in two-photon 
exchange could have been significantly underestimated. More recently, Carlson and 
Vanderhaeghen \cite{cv11b} have examined the constraints on $T_{1,2}(\omega,Q^2)$ 
from real Compton scattering and shown that these are incompatible with the 
model of Miller \textit{et al.}~\cite{mtcr11}.

To obtain a model-independent result for the form of $T_1(0,Q^2)$ at small $Q^2$,
we calculate $T_1(0,Q^2)$ within the framework of chiral perturbation theory (ChPT) 
to fourth order. The polarisability contribution to the Lamb shift has previously 
been calculated using heavy-baryon chiral perturbation theory (HBChPT) at third order 
and found to be small \cite{np08}. However it is known that fourth-order contributions 
to the polarisabilities of the proton are larger than naively expected \cite{bksm93}. 
In addition, the $\Delta$ resonance is known to make an important contribution to the 
magnetic polarisability $\beta$ \cite{bs92,hhk97,pp03}, and its potential 
significance for the Lamb shift has been stressed by Pineda \cite{pin05}. 
We also note that ChPT results for spin-dependent V$^2$CS 
show rapid dependence on $Q^2$ at low momenta \cite{bkm93,jko00,bhm02}. If the 
spin-independent amplitude $T_1(0,Q^2)$ were to show similar behaviour, this could be
important for the Lamb shift. These results imply that it is important to 
check whether higher-order effects could alter previous conclusions. We do so here
using the same ChPT approach as recently applied to the analysis of real Compton
scattering \cite{gmpf12}, treating nucleons to fourth-order in HBChPT and including
effects of the $\Delta$ up to fifth order in the ``$\delta$-counting" introduced 
by Pascalutsa and Phillips \cite{pp03}. This gives us an estimate of the ``mass" in
the form factor for $T_1(0,Q^2)$ which is determined consistently with the value 
for $\beta$ obtained in Ref.~\cite{gmpf12}.

In view of the questions that have been raised about the LET for $T_1(0,Q^2)$, 
we re-examine its derivation to clarify some of the issues related to it and
to the separation of the amplitude into Born and ``structure" parts.
We also comment on how the LET is embodied in nonrelativistic effective field 
theories such as HBChPT.

Finally, given the low-energy nature of the theory, it cannot give reliable 
estimates of the behaviour for larger momenta but we give arguments in
favour of a $1/Q^2$ fall-off at large $Q^2$, as also obtained
from the operator-product expansion \cite{col79}. We use this to estimate the 
contribution to the Lamb shift and to show that it is too small to explain the 
deduced discrepancy.
 
\section{Low-energy limit}

Before calculating the higher-order contributions of proton structure to $T_1$, 
we first re-examine the low-energy constraints on the proton-structure parts of 
the amplitude. In the low-momentum limit, the amplitudes 
for real and virtual Compton scattering can be related through the limited number 
of tensor structures that respect covariance and gauge invariance, and are even 
under crossing ($q\leftrightarrow -q'$, $\nu\leftrightarrow\mu$). The invariant
amplitudes multiplying them should also be free from unphysical kinematic 
singularities \cite{bt68,tar75} (see also: Refs.~\cite{dkms97,fs98}). Under 
these constraints, the spin-independent amplitude can be expanded in terms of 
five basis tensors:
\begin{equation}
T^{\mu\nu}=\sum_{i=a,\cdots,e}T_i(\bar p^2,q^2,q\cdot \bar p,q'\cdot q)\,
t_i^{\mu\nu},
\end{equation}
where $\bar p=(p+p')/2$ is the average of the initial and final proton momenta, 
and we have taken the initial and final photons to have the same
virtualities.\footnote{If the virtualities are different, there is also a 
sixth basis tensor.} These tensors may be chosen to be:
\begin{eqnarray}
t_a^{\mu\nu}&=&-q'\cdot q\,g^{\mu\nu}+q^\mu q^{\prime\nu},\cr
\noalign{\vspace{5pt}}
t_b^{\mu\nu}&=& q'\cdot q\,\bar p^\mu \bar p^\nu-q\cdot \bar p\,
(\bar p^\mu q^{\prime\nu}+q^\mu \bar p^\nu)+(q\cdot \bar p)^2\,g^{\mu\nu},\cr
\noalign{\vspace{5pt}}
t_c^{\mu\nu}&=&q^4 g^{\mu\nu}+q'\cdot q\;q^{\prime\mu}q^\nu
-q^2(q^\mu q^\nu+q^{\prime\mu}q^{\prime\nu}),\cr
\noalign{\vspace{5pt}}
t_d^{\mu\nu}&=&q\cdot \bar p\!\left[2\,q\cdot \bar p\;q^2g^{\mu\nu}
-q\cdot \bar p\,(q^\mu q^\nu+q^{\prime\mu}q^{\prime\nu})\right.\cr
\noalign{\vspace{5pt}}
&&\qquad\left.-q^2(\bar p^\mu q^{\prime\nu}+q^\mu \bar p^\nu)
+q'\cdot q(\bar p^\mu q^\nu+q^{\prime\mu}\bar p^\nu)\right],\cr
\noalign{\vspace{5pt}}
t_e^{\mu\nu}&=&q^4\bar p^\mu \bar p^\nu+(q\cdot \bar p)^2q^{\prime\mu}q^\nu
-q\cdot \bar p\;q^2(\bar p^\mu q^\nu+q^{\prime\mu}\bar p^\nu).\cr
\noalign{\vspace{5pt}}
&&\label{eq:tensors}
\end{eqnarray}
Note that only four of the tensors are linearly independent, 
but there is no way to eliminate any one of them without introducing 
kinematic singularities in the amplitudes $T_i$ \cite{tar75,dkms97,fs98}.

In calculating the two-photon-exchange interaction between a lepton and a proton,
it is convenient to split the Compton amplitude into a Born piece, expressed 
entirely in terms of couplings of a single photon to an on-shell proton, and a 
``structure" piece involving, for example, the polarisabilities of the proton.
Conventionally, the Born amplitude is calculated from a Dirac equation for 
the proton with Dirac and Pauli form factors, $F_D(Q^2)$ and $F_P(Q^2)$ 
respectively (often denoted $F_1$ and $F_2$ although this invites confusion 
with inelastic structure functions).
The Born contributions to the spin-independent Compton amplitude\footnote{This 
has been defined by taking a covariant spin average, 
$T^{\mu\nu}={\rm Tr}[({p\llap/}^\prime+M)T_s^{\mu\nu}({p\llap/}+M)]/(8M^2)$,
similar to that used in Ref.~\cite{bt76}. Up to an overall factor, 
it corresponds to the spin-independent amplitude in the Breit frame, where there 
is a unique spin axis defined by the (parallel) initial and final momenta of 
the proton.} 
obtained in this way have the forms
\begin{eqnarray}
T^B_a&=&e^2\,\frac{\left(\bar p^2q'\cdot q-(q\cdot\bar p)^2\right)
F_P\left(2F_D+F_P\right)}{M^3(s-M^2)(u-M^2)},\cr
\noalign{\vspace{5pt}}
T^B_b&=&e^2\,\frac{q'\cdot q\,F_P\left(2F_D+F_P\right)
-4M^2F_D^2-2q^2F_DF_P}{M^3(s-M^2)(u-M^2)},\cr
\noalign{\vspace{5pt}}
T^B_c&=&0,\cr
\noalign{\vspace{5pt}}
T^B_d&=&-e^2\,\frac{F_P\left(F_D+F_P\right)}{M^3(s-M^2)(u-M^2)},\cr
\noalign{\vspace{5pt}}
T^B_e&=&0.
\end{eqnarray}
As required for the amplitudes corresponding to the tensors of 
Eq.~(\ref{eq:tensors}), these have poles only at the on-shell points, 
$s\equiv(p+q)^2=M^2$ and $u\equiv(p-q')^2=M^2$.

The standard treatment of the Lamb shift \cite{pach99} subtracts this Born 
amplitude to leave a residual, structure amplitude that is free of nucleon poles 
and can be estimated from inclusive-scattering structure functions with the aid 
of dispersion relations. This approach has been dismissed by Hill and Paz 
\cite{hp11} as ``sticking in form factors". However it provides a well-defined
expression for the $Q^2$ dependence of the Born amplitude and, provided that 
the non-Born piece is defined consistently with it, there is no problem with 
this choice. Below, we comment further on the need for consistency 
between the choice of Born amplitude and the form of the non-Born piece.

First, we look at the LETs satisfied by the non-Born amplitude defined in this
way. This amplitude starts at second order in the photon momenta,
with terms proportional to $t_a$ and $t_b$, the only tensors of order $q^2$:
\begin{equation}
T^{\mu\nu}=T^{B\mu\nu}-4\pi\beta\,t_a^{\mu\nu}
-\frac{4\pi}{M^2}(\alpha+\beta)\,t_b^{\mu\nu}+{\cal O}(q^4),
\label{eq:csq2}
\end{equation}
where the proton states are normalised so that the Thomson term is 
$(e^2/M)\epsilon'\cdot\epsilon$.
The contributions of proton structure at this order are contained in the two
constants multiplying these tensors. They have been expressed in
terms of the electric and magnetic polarisabilities of the proton, $\alpha$ and 
$\beta$, which can be determined from real Compton scattering at low energies.
(For details, see the review: Ref.~\cite{gmpf12}.) 

In the case of \emph{forward} V$^2$CS, the standard amplitudes in 
Eq.~(\ref{eq:fv2cs}) can be related to those multiplying the five 
tensors of Eq.~(\ref{eq:tensors}) by
\begin{eqnarray}
T_1&=&q^2T_a-(q\cdot p)^2T_b-q^4T_c-2q^2(q\cdot p)^2T_d,\cr
\noalign{\vspace{5pt}}
T_2&=&M^2(q^2T_b+q^4T_e).
\end{eqnarray}
The residual amplitudes $\overline T_i=T_i-T_i^B$ satisfy the LETs 
\cite{bt76},
\begin{eqnarray}
\overline T_1(\omega,Q^2)&=&4\pi\,Q^2\beta+4\pi\,\omega^2(\alpha+\beta)
+{\cal O}(q^4),\label{eq:let1}\\
\noalign{\vspace{5pt}}
\overline T_2(\omega,Q^2)&=&4\pi\,Q^2(\alpha+\beta)+{\cal O}(q^4).
\label{eq:let2}
\end{eqnarray}
The LET for $T_1(0,Q^2)$ is important in the standard approach to the Lamb 
shift \cite{pach99} since the dispersion relation for $\overline T_1$
requires a subtraction and the LET constrains the low-$Q^2$ limit of the 
subtraction term.

Any introduction of off-shell form factors along the lines suggested 
by Miller et al.~\cite{mtcr11} should respect these LETs. The fact that 
their results for $T_1(0,Q^2)$ do not is more than just a 
numerical inconsistency with real Compton scattering \cite{cv11b}, 
it represents a violation of general principles.

We now return to the question of consistency between Born and non-Born amplitudes.
The factors of the Breit-frame photon energy $q\cdot \bar p$ in the basis tensors 
of Eq.~(\ref{eq:tensors}) mean that the Born contributions to the full Compton 
amplitude $T^{\mu\nu}$ contain pieces without poles corresponding to on-shell 
intermediate proton states. This can be seen clearly in the case of forward
 V$^2$CS, where the Born terms have the forms
\begin{eqnarray}
T_1^B(\omega,Q^2)&=&\frac{e^2}{M}\Biggl[\frac{Q^4\left(F_D(Q^2)+F_P(Q^2)\right)^2}
{Q^4-4M^2\omega^2}-F_D(Q^2)^2\Biggr],\label{eq:T1Born}\cr
\noalign{\vspace{5pt}}
&&\qquad\qquad\label{eq:t1b}\\
\noalign{\vspace{5pt}}
T_2^B(\omega,Q^2)&=&\frac{4e^2MQ^2}{Q^4-4M^2\omega^2}\cr
\noalign{\vspace{5pt}}
&&\times\left[F_D(Q^2)^2+(Q^2/4M^2)F_P(Q^2)^2\right],
\end{eqnarray}
and the second piece of $T^B_1$ contains no poles at $\omega=\pm Q^2/2M$. 

Various authors (for recent examples, see Refs.~\cite{cv11a,hp11,wlcm12}) have 
suggested that, rather than splitting the Compton amplitude into ``Born" and 
``structure" pieces, it is better separate it into ``pole" and ``non-pole" pieces. 
In that approach, the second term of Eq.~(\ref{eq:T1Born}) would count as part of 
the non-pole amplitude, along with the polarisabilities and other structure 
contributions. However, as shown by Walker-Loud \textit{et al.}~\cite{wlcm12}, 
this separation is not unique since the form of the pole piece depends on the 
choice of basis tensors. In the basis of Eq.~(\ref{eq:tensors}) above, the 
$F_D(Q^2)^2$ term arises from a pole term in the amplitude $T_b$ and,
furthermore, it is required to keep the full Born amplitude free from kinematic 
singularities at $Q^2=0$. All of this implies that the separation into Born
plus structure is the more natural one.

The role of this term can be seen more clearly by expanding it in powers of 
$Q^2$ to get
\begin{equation}
F_D(Q^2)^2=1-\left[\frac{1}{3}\,r_{c}^2-\frac{\kappa}{2M^2}\right]Q^2
+{\cal O}(Q^4),
\label{eq:fd2}
\end{equation}
where $r_c$ is the charge radius of the proton (defined in terms of the slope 
of the Sachs electric form factor) and $\kappa=F_P(0)$ is its anomalous 
magnetic moment. This shows that the leading term in this 
piece of $T^B_1$ gives rise to a term proportional to $q_\mu q_\nu/q^2$ in the 
scattering amplitude. In the full amplitude this cancels against a similar 
singular term arising from $T^B_2$.

The next term in $F_D(Q^2)^2$, of order $Q^2$, generates a contribution to the 
scattering amplitude that remains finite for real photons, $Q^2\rightarrow 0$. 
Shifting this contribution from the Born amplitude to the structure part
would alter the coefficient of $Q^2$ in $\overline T_1$ and so would correspond 
to a definition of the magnetic polarisability $\beta$ that differed from 
the conventional one. Carlson and Vanderhaeghen \cite{cv11a} keep only the pole 
pieces of the Born amplitude but the form they take for $\overline T_1(0,Q^2)$ 
is inconsistent with this since it satisfies an LET with conventional definition 
of $\beta$. The fact that the order-$Q^2$ term in $T_1(0,Q^2)$ contains this 
piece from the expansion of $F_D(Q^2)^2$ leads Walker-Loud 
\textit{et al.}~\cite{wlcm12} to use the same definition for the structure part 
of the V$^2$CS amplitude as we advocate, albeit without the explicit
justification presented here.

In the context of the HBChPT calculation described in the next section, 
consistency between the Born and structure amplitudes as usually defined means 
that we must subtract the full Born amplitude of Eq.~(\ref{eq:t1b}) expanded to 
the appropriate order in $Q^2$. Specifically, in addition to all pole terms,
we need to subtract the terms in the expansion of $F_D(Q^2)^2$. To do this we 
need the Dirac form factor calculated to third order in the chiral expansion, 
which is given in Ref.~\cite{bkm95}.

The LETs of Eqs.~(\ref{eq:let1}, \ref{eq:let2}) are built into the effective field
theory for a nonrelativistic particle through the constraints of relativity, often
expressed in terms of ``reparametrisation invariance" \cite{lm92}. The Lagrangian 
of any such theory contain three types of structure with two derivatives of the 
photon fields \cite{man97,fmms00}. Two of these contribute to the polarisabilities 
$\alpha+\beta$ and $\beta$. In contrast, the third has a coefficient that is fixed 
in terms of lower-order constants that correspond to the charge radius and anomalous
magnetic moment. It generates the order-$Q^2$ term of the non-pole piece of $T^B_1$
discussed above. This is illustrated by the amplitude calculated in nonrelativistic
QED, Eq.~(11) of Ref.~\cite{hp11}, which contains both the non-pole Born piece 
arising from the second term of Eq.~(\ref{eq:fd2}) and the magnetic polarisability.

\section{Chiral perturbation theory}

The specific amplitude we consider here is $T_1(0,Q^2)$ for forward V$^2$CS, which 
appears as the subtraction term in the dispersion relation for $T_1(\omega,Q^2)$ 
\cite{pach99} and so forms one of the least-well determined contributions of proton
structure to the Lamb shift \cite{mtcr11,cv11a,hp11}. Its leading term, of order 
$Q^2$, is fixed by the LET discussed in the previous section but contributions from 
its higher-order terms vanish in the real-photon limit and so are subject to no 
such constraints. If these terms were to grow rapidly with $Q^2$, they could lead 
to unexpectedly large structure contributions to the Lamb shift. They have been 
calculated by Nevado and Pineda \cite{np08} at third-order where, like the
polarisabilities \cite{bkm92,bkm95}, they are predictions that do not require any 
two-photon low-energy constants. However there are large contributions at 
fourth-order to both spin-independent and spin-dependent real Compton scattering 
\cite{bksm93,bkm93,jko00,bhm02}. In addition, the effects of $\Delta$ resonance 
are important for the magnetic polarisability and could also play a significant
role in the Lamb shift \cite{pin05}. To examine whether these effects 
could be similarly important for the Lamb shift, we have calculated the $Q^4$ 
piece of forward V$^2$CS treating nucleons to fourth order in HBChPT. We also 
include the leading contribution of the $\gamma$N$\Delta$ form factor, a term of
fifth order in the $\delta$ counting of Ref.~\cite{pp03}.

The calculation of $T_1(0,Q^2)$ has a very close correspondence to that of forward 
real Compton scattering \cite{bkm92,bkm95,mcg01}. To take advantage of this, 
we choose a gauge in which the polarisation vectors of the virtual photons are 
purely space-like (in the proton rest frame), so that the conditions 
$p\cdot\epsilon =0$, $q\cdot\epsilon =0$ which were used in the real case still 
hold. With this choice, the diagrams that contribute are the same, and in many 
cases the resulting expressions can be obtained from the corresponding ones for 
real photons with the aid of some straightfoward substitutions of kinematic 
variables. A representative subset of the fourth-order nucleon diagrams is shown 
in Fig.~1; the full set is given in Fig.~1 of Ref.~\cite{mcg01}. (The third-order
diagrams can be found in Fig.~1 of Ref.~\cite{np08}.)

\begin{figure*}[t]
\begin{center}\includegraphics*[width=0.85\hsize]{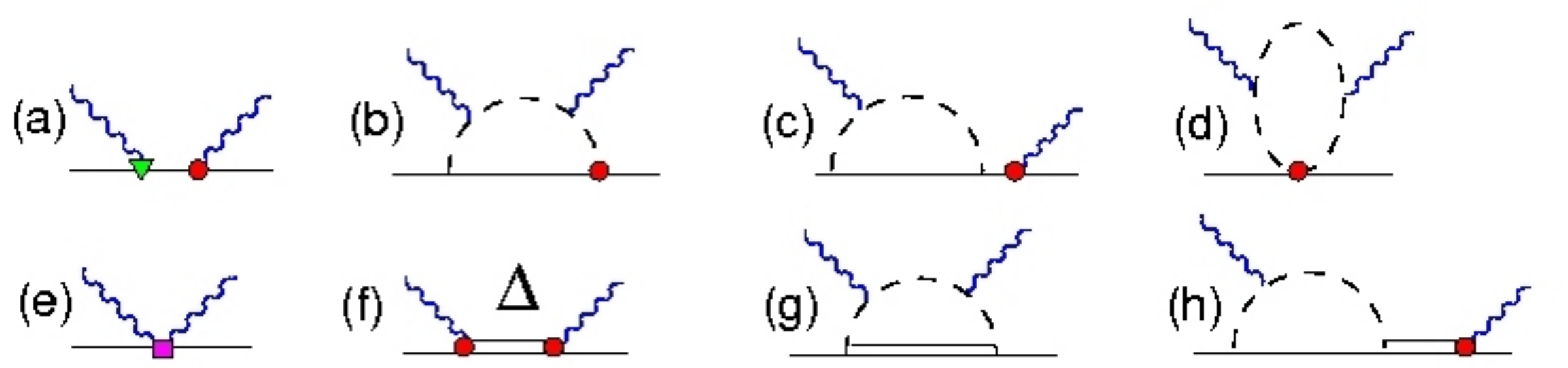}
\caption{Some diagrams contributing to Compton scattering in 
HBChPT: (a) a one-nucleon-reducible tree graph;
(b--d) a representative selection of $\pi$N loop graphs that contribute at 
fourth order; (e) the fourth-order seagull contribution to $\beta$;
(f, g) $\Delta$ contributions at order $\delta^3$ and (h) at $\delta^5$.
In these graphs the circle, triangle and square denote vertices from the second-,
third- and fourth-order Lagrangians, respectively.}
\end{center}
\end{figure*}

To determine the ``structure" piece of the amplitude, $\overline T_1$, we need to
separate it from the Born terms. This separation is somewhat subtle in HBChPT, and 
cannot just be done on a diagram by diagram basis \cite{mcg01}. Some diagrams, 
such as Fig.~1(a), can immediately be identified as parts of the ``magnetic" 
Born term (the first, ``pole" term of Eq.~(\ref{eq:t1b})). 
However there is also a one-nucleon-reducible diagram in which 
one photon couples to the nucleon via a magnetic-moment coupling and the other via 
a pion loop, Fig.~1(c). As discussed in Ref.~\cite{bjm01}, 
this contributes to both the magnetic Born and structure pieces of real Compton 
scattering. The structure part arises from the energy dependence of the pion loop 
which corresponds to an ``off-shell form factor" of the sort proposed by Miller 
\textit{et al.}~\cite{mtcr11}, but with a calculable form. This illustrates the 
fact that such off-shell dependences cannot be distinguished from structure effects
such as polarisabilities.

In the case of space-like virtual photons considered here, particular care is needed
with the one-nucleon reducible diagrams since the the leading propagator for the
intermediate nucleon is $2M/Q^2$. This means that such diagrams contribute at one 
order lower than in the expansion of real Compton scattering. For example, the 
diagram Fig.~1(c) just discussed contributes at third order, where it forms part 
of the magnetic Born piece of $T_1$. At fourth order there is a very similar 
graph but with a second-order insertion in the nucleon propagator within the loop. 
This contributes to both Born and ``structure" pieces of the amplitude. To this 
order, all other reducible diagrams with the enhanced $2M/Q^2$ propagator (diagrams
like Fig.~1(a) but with higher-order vertices) contribute only to the magnetic Born
term.

The sum of all graphs (at third and fourth orders) has no constant term; the 
coefficient of $Q^2$ is the same as the loop contribution to $\beta$ 
\cite{bksm93} plus a piece proportional to the third-order loop contribution to 
the derivative of the single-photon form-factor $F_D$. The divergences in these
loop integrals are renormalised by photon-nucleon sea\-gulls, Fig.~1(e), arising 
from terms in the fourth-order chiral Lagrangian. These include a term with the
structure $F_{\mu\nu}F^{\mu\nu}$ which is the same one that appears in the 
renormalisation of the magnetic polarisability $\beta$. This term is a combination 
of $\widehat O^{(4)}_{89,91,93,118}$ in the minimal Lagrangian of Fettes 
\textit{et al.}~\cite{fmms00}. That Lagrangian also contains terms with the 
structure $v^\mu v^\nu F_{\lambda\mu}F^\lambda_{\;\nu}$ which contribute to 
$\alpha+\beta$ and so are not relevant to forward V$^2$CS. 

In addition, the fourth-order heavy-baryon effective Lagrangian contains 
terms with coefficients that are fixed by relativistic invariance in terms 
of coefficients from the lower-order Lagrangians. A combination of 
two of these is needed here, $X_{40}$ and $X_{52}$ of Ref.~\cite{fmms00}. 
These have the structure $[D_\mu,F^{\mu\lambda}]D_\lambda+{\rm h.c.}$, which
does not correspond to either of the polarisabilities of the nucleon. But, 
like the similar $c_M$ term in the norelativistic QED Lagrangian of Manohar 
\cite{man97}, the coefficient of this is given in terms of the second-order 
anomalous magnetic moment of the proton and the third-order coefficient that 
contributes to the charge radius. In the present context, this seagull diagram
renormalises the loop contributions of order $Q^2$ to $F_D(Q^2)^2$, the 
``non-pole" Born term. 

As well as fourth-order terms in the chiral expansion, we have also examined 
the effect of explicitly including the $\Delta$ resonance since this low-lying 
excitation can play an important role in the responses of a nucleon to external 
fields, particularly in the magnetic polarisability \cite{bs92,hhk97,pp03}.
Working within the $\delta$ expansion introduced by Pascalutsa and Phillips 
\cite{pp03}, we find that the leading (order-$\delta^4$) $\Delta$-pole 
graph, Fig.~1(f), has the same form as the seagull contribution to $\beta$ and 
so can simply be absorbed in the corresponding low-energy constant. 
In a similar way, the leading $\pi\Delta$ loops such as Fig.~1(g) can to a 
good approximation be absorbed into the the values of the low-energy constants 
appearing in the fourth-order tadpole diagrams, Fig.~1(d). The full expression 
differs from the pieces already included in the fourth-order result by about 
16\%, which is well within the uncertainty introduced by the rather poorly-known 
second-order constants, and so we do not give it explicitly.

The one $\Delta$-pole contribution we do include arises from Fig.~1(h), where the 
$\pi$N loop generates a $\gamma$N$\Delta$ form factor. This diagram is of fifth 
order in $\delta$, which is higher than the other terms considered here, but it
provides the leading contribution of this form factor to $T_1(0,Q^2)$ at order 
$Q^4$. We have therefore included it as an estimate of the likely influence of the 
$\Delta$.

After subtracting $T^B_1$ as discussed above, we arrive at the following result for 
$\overline T_1$:
\begin{eqnarray}
&&\overline T_1(0,Q^2)\nonumber\\
\noalign{\vspace{5pt}}
&&=4\pi\beta\, Q^2-\biggl[\frac{e^2g_{\scriptscriptstyle A}^2}
{1280\pi f_\pi^2\, m_\pi^3}
\nonumber\\
\noalign{\vspace{5pt}}
&&\qquad+\frac{e^2}{480\pi^2f_\pi^2\,m_\pi^2}\biggl(4c_1
+c_2-2c_3-\frac{g_{\scriptscriptstyle A}^2}{M_{\scriptscriptstyle\rm N}}
(4+5\mu_s)\biggr)\nonumber\\
\noalign{\vspace{5pt}}
&&\qquad+\frac{e^2g_{\scriptscriptstyle M}\,g_{\scriptscriptstyle A}\,
g_{\pi{\scriptscriptstyle {\rm N}\Delta}}}{144\pi f_\pi^2\,m_\pi 
(M_{\scriptscriptstyle \Delta}^2-M_{\scriptscriptstyle\rm N}^2)}
\biggr]Q^4+{\cal O}(Q^6).
\label{eq:t1str}
\end{eqnarray}
Full expressions for the contributions to the amplitude, not 
expanded in powers of $Q^2$, are given in the Appendix.

In evaluating this numerically, we take $g_{\scriptscriptstyle A}=1.27$,
$f_\pi=92.2$~MeV, and $\mu_s=0.88$. The constants $c_i$ from the second-order
Lagrangian have been determined from $\pi$N scattering and the resulting values, 
as quoted by Bernard \cite{ber07}, are $c_1=-0.9^{+0.2}_{-0.5}$, $c_2=3.3 \pm 0.2$, 
$ c_3=-4.7^{+1.2}_{-1.0}$, all in GeV$^{-1}$. However a more tightly constrained 
value of $c_3=-4.78\pm0.2$ GeV$^{-1}$ has been obtained from $pp$ scatttering 
\cite{rts03}, and we use this here. 

For the couplings of the $\Delta$, we 
take the values obtained by Pascalutsa \textit{et al.}~\cite{pv05,pvy06} of
$g_{\scriptscriptstyle M}=2.9$ for the magnetic $\gamma$N$\Delta$ coupling 
constant and $g_{\pi{\scriptscriptstyle {\rm N}\Delta}}=1.425$. These couplings 
provide a good description of real Compton scattering below the $\Delta$ peak
\cite{gmpf12} but they do depend on the choice of Lagrangian. Nonetheless a very
similar value for the product $g_{\scriptscriptstyle A}\,
g_{\pi{\scriptscriptstyle {\rm N}\Delta}}$ has also been found to give a good 
description of Compton scattering within a purely nonrelativistic framework for the  
$\Delta$ \cite{hghp04}, despite $\sim30\%$ differences in the separate couplings.
A more serious source of uncertainty is the fact that these analyses are mainly
sensitive to the couplings close to the $\Delta$ peak, whereas we need the
values at zero energy. To account for this we allow for a 20\% error on 
$g_{\scriptscriptstyle A}\, g_{\pi{\scriptscriptstyle {\rm N}\Delta}}$

If we express our result in the form of a 
form factor, 
\begin{equation}
\overline T_1(0,Q^2)\simeq 4\pi\beta\, Q^2\left(1-\frac{Q^2}{M_\beta^2}
+{\cal O}(Q^4)\right),
\end{equation}
its slope is given by
\begin{equation}
M_\beta^2=\frac{\beta}{3 \times 10^{-4}\text{fm}^3}
\,(455\pm 32\; \text{MeV})^2.
\label{eq:mbeta}
\end{equation}

The magnetic polarisability $\beta$ has recently been determined from a fit
to real Compton scattering within the same framework as used here, giving 
$\beta=(3.1\pm0.5)\times 10^{-4}$~fm$^3$ (including the statistical and Baldin 
Sum Rule errors only) \cite{gmpf12}. Using this and the values for the $c_i$ 
and $\Delta$ couplings above, with their errors, gives $M_\beta=460\pm 50$~MeV 
for the mass parameter in $\overline T_1(0,Q^2)$. This has a size $\sim 3m_\pi$, 
which is ``natural" for an effect with important contributions from pion loops.

Without the fourth-order and $\Delta$ contributions to the slope, 455~MeV would 
be replaced by 588~MeV in Eq.~(\ref{eq:mbeta}). In view of this change from third 
to fourth order, the residual error due to neglect of higher-order terms is 
expected to be comparable to the uncertainty arising from the low-energy 
constants. However the $\Delta$ contributions are likely to provide a larger 
source of uncertainty as the expansion parameter in the $\delta$ expansion is 
not particularly small. If we triple the error associated with of the order-$Q^4$ 
term in Eq.~(\ref{eq:t1str}) to take account of these uncertainties, then we get 
\begin{equation}
M_\beta=460\pm 100 \pm 40\; \text{MeV},
\end{equation}
where the first error is from the order-$Q^4$ term and the second is due to $\beta$.

\section{Lamb shift}

The calculation outlined in the previous section leads to a model-independent 
result for the V$^2$CS amplitude $\overline T_1(0,Q^2)$ up to order $Q^4$.
This constrains the subtraction term in the dispersion relation for 
$\overline T_1(\omega,Q^2)$ in a way that does not rely on the \textit{ad hoc}
use of the magnetic form factor of the proton as in Ref.~\cite{pach99}. 
However it is not sufficient to estimate the polarisability contribution to 
the muonic Lamb shift, since that requires an integral over $Q^2$. 

Our chiral calculation gives an expression for $\overline T_1(0,Q^2)$ as a 
function of $Q^2$ which can be found in the Appendix. However, this cannot be 
used for large $Q^2$ since it contains terms (such as the fourth-order seagull)
that do not vanish as $Q^2\rightarrow \infty$. Such terms were not present in 
the third-order calculation of Nevado and Pineda \cite{np08}, where the only 
photon-nucleon seagull is the one that gives the Thomson limit and so is part 
of the Born amplitude.

Other information on the form of $T_1(0,Q^2)$ comes from the large-$Q^2$ limit 
where the operator product expansion leads to $1/Q^2$ for the dominant 
behaviour \cite{col79} (as noted in Ref.~\cite{hp11}). This form can also 
be understood from the quark counting rules that apply to the partonic regime
\cite{vz78,cg84}. In this regime there is no single-quark contribution to 
$T_1(0,Q^2)$, as can be seen from the exact cancellation of the two terms in 
Eq.~(\ref{eq:T1Born}) at $\omega=0$ for pointlike fermions. Instead one-gluon 
exchange between two quarks gives the dominant contribution and the counting rules 
show that this falls off like $1/Q^2$. 

Here we estimate the contribution to the Lamb shift by matching our result 
smoothly onto the expected high-$Q^2$ behaviour. To do this we write
\begin{equation}
\overline T_1(0,Q^2)=4\pi\beta\,Q^2\,F_\beta(Q^2),
\end{equation}
and take the dipole form 
\begin{equation}
F_\beta(Q^2)=\frac{1}{\left(1+Q^2/2\overline M_\beta^2\right)^2}\,,
\label{eq:extrap}
\end{equation}
for the form factor. 
This matching could be carried out at $Q^2=0$, in which case $\overline M_\beta$
is just the value $M_\beta$ determined from the slope of the ChPT form factor 
in the previous section. However nonanalytic terms arising from pion loops
can lead to non-dipole dependences of form factors on $Q^2$ in the 
region $Q^2\lesssim (3m_\pi)^2$ \cite{bkkm92,bfhm98}. In Fig.~2 we show the 
full ChPT result for the form factor $F_\beta(Q^2)$
compared to dipole forms with masses $\overline M_\beta=460$ and 510~MeV. 
We see that the full form factor and the extrapolation from $Q^2=0$ do deviate 
for $Q^2\gtrsim 0.03$~GeV$^2$, although the differences lie well within the
uncertainties of our calculation. In fact the ChPT result in the region from  
$Q^2\simeq 0.1$ to nearly $0.3$~GeV$^2$ has a form that is very similar to a 
dipole with a mass of about 510~MeV. If we match onto the dipole form anywhere 
in this region we get a mass close to this value. In estimating the 
contribution to the Lamb shift we have therefore used Eq.~(\ref{eq:extrap}) with 
\begin{equation}
\overline M_\beta=485\pm 100\pm 40\pm 25\; \text{MeV},
\end{equation}
where the first two errors arise from the order-$Q^4$ term and $\beta$, as discussed
above, and the last is from the choice of matching point for the extrapolation. 
The shaded region in Fig.~2 corresponds to our estimate of the the uncertainties 
associated with our calculation of the order-$Q^4$ term by showing dipole forms 
with masses in the range 385 to 585~MeV.

\begin{figure}[ht]
\begin{center}\includegraphics[width=8.5cm]{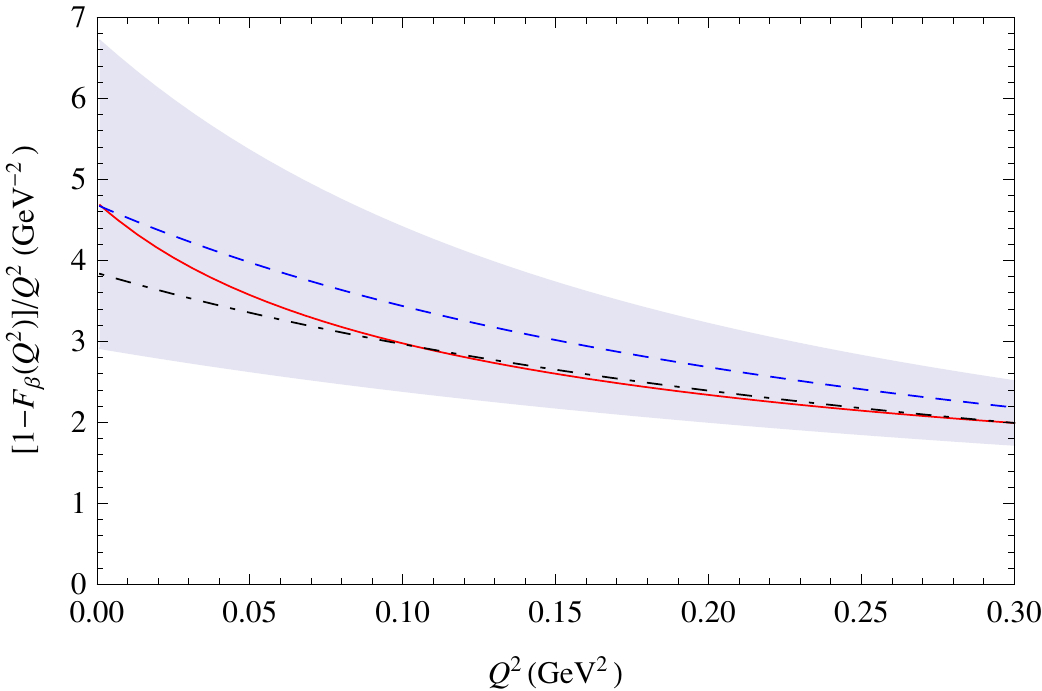}
\caption{The slope of the form factor for V$^2$CS, plotted in the form
$[1-F_\beta(Q^2)]/Q^2$ to emphasise variations at small $Q^2$. 
The solid (red) curve shows the full result of ChPT calculation. The dashed 
(blue) curve is the dipole form of Eq.~(\ref{eq:extrap}) fitted to the slope 
at $Q^2=0$. The dot-dashed (black) curve is a dipole with a mass 
$\overline M_\beta=510$~MeV. The shaded region shows dipole forms with masses 
in range 385 to 585~MeV.}
\end{center}
\end{figure}

The contribution of the subtraction term to the Lamb shift is, from Eq.~(34) of 
Ref.~\cite{pach99},\footnote{Note that our definition of $T_1$ differs from that 
of Ref.~\cite{pach99} by a factor of $e^2M$.}
\begin{eqnarray}
\Delta E_{\rm sub}&=&\frac{\alpha_{\scriptscriptstyle\rm EM}\,\phi(0)^2}
{4\pi\, m}\int_0^\infty{\rm d}Q^2\,\frac{\overline T_1(0,Q^2)}{Q^2}
\nonumber\\
\noalign{\vspace{5pt}}
&&\qquad\times\left[1+\left(1-\frac{Q^2}{2m^2}\right)
\left(\sqrt{\frac{4m^2}{Q^2}+1}-1\right)\right],\nonumber\\
\end{eqnarray}
where $m$ is the lepton mass and 
\begin{equation}
\phi(0)^2=\frac{1}{8\pi}\left(\frac{\alpha_{\scriptscriptstyle\rm EM}\,m\,M_p}
{m+M_p}\right)^3
\end{equation} 
is the square of the $2S$ wave function at the origin. Although we evaluate this
taking the dipole form of Eq.~(\ref{eq:extrap}) for all $Q^2$, we have also
checked that its integral for $\overline M_\beta \simeq 510$~MeV agrees to
better than 1\% with that of the full form matched onto a dipole at 
$Q^2=0.2$~GeV$^2$. 
Note that the leptonic factor strongly weights the low-$Q^2$ region of this 
integral, with 90\% coming from the region $Q^2\leq 0.3$~GeV$^2$. This means 
that matching onto the correct form for larger values of $Q^2$ is not crucial, 
provided that the assumed form does not have a much longer tail than the dipole. 

Evaluating the integral using Eq.~(\ref{eq:extrap}) for $\overline T_1(0,Q^2)$ 
we get
\begin{equation}
\Delta E_{\rm sub}=4.2\pm 1.0\;\mu\mbox{eV},
\end{equation}
for the contribution to the Lamb shift in muonic hydrogen.
The error here includes our estimate of the uncertainties associated with 
higher-order effects, as discussed in the previous section. However
we note that largest single uncertainty comes from the value for $\beta$,
which appears both as an overall factor in $\overline T_1(0,Q^2)$ 
and in the mass parameter $M_\beta$, as given by Eq.~(\ref{eq:mbeta}).

Our result for $\Delta E_{\rm sub}$ is more than twice that obtained by 
Pachucki \cite{pach99}. Most of the difference is due to the modern value for 
$\beta$, which is about twice the one used in that work. There is also a further 
$\sim 10$\% increase from the use of the form factor with the correct asymptotic 
form. Our result is smaller that the one of Carlson and Vanderhaeghen \cite{cv11a} 
who used a similar value for $\beta$ as ours but whose model form factor leads to 
$T_1(0,Q^2)$ growing like $\ln Q^2$ asymptotically.

Finally, to estimate the total two-photon exchange contribution to the Lamb shift, 
we add in the ``elastic" (Born) and dispersive contributions from Pachucki's 
treatment \cite{pach96,pach99}. We use the updated numerical results from Carlson 
and Vanderhaeghen \cite{cv11a} and, for consistency with the definition of $\beta$, 
we reinstate the ``non-pole" Born piece. The latter gives about 4.8~$\mu$eV 
for the form factor of Ref.~\cite{amt07} and so is comparable in importance to the 
subtraction term. The main sources of uncertainty are $\beta$ in our subtraction 
term above and the dependence of the elastic contribution on the choice of 
parametrisation of the proton form factors (which we take as $\pm 1.6$~$\mu$eV,
based on the results in Ref.~\cite{cv11a}). The bottom line for the energy is
\begin{equation}
\Delta E_{2\gamma}=-33\pm 2\;\mu\mbox{eV},
\end{equation}
which is not significantly different from the values of about $-35$~$\mu$eV obtained 
in Refs.~\cite{pach99,mart06} but which does not rely on a model for the form factor
in $\overline T_1(0,Q^2)$.

\section{Summary}

In this work, we have calculated the V$^2$CS amplitude $T_1(0,Q^2)$ at fourth order 
in HBChPT, and including contributions of the $\Delta$ resonance. Our result 
provides a model-independent constraint on the low-momentum form of the structure 
part of this amplitude. Since the contribution of this amplitude to the Lamb
shift is dominated by the low-momentum region, this can significantly 
reduce the theoretical uncertainty in contribution of the polarisability of the 
proton to the Lamb shift, which has been the subject of 
some recent debate \cite{mtcr11,cv11a,hp11,cv11b}.

When our amplitude is matched smoothly onto the high-$Q^2$ behaviour expected in
the partonic regime \cite{hp11}, we obtain a contribution to the Lamb shift that,
perhaps unsurprisingly, is similar in magnitude to previous, more model-dependent 
determinations \cite{pach99,cv11a}.

In reaching our result we have re-examined the LETs for V$^2$CS and how they 
are embodied in effective field theories. The constraints of covariance, 
gauge invariance and crossing symmetry as well as lack of kinematic 
singularities mean that the only structure terms of order $Q^2$ are given
by the electric and magnetic polarisabilities. This is reflected by the 
appearance in fourth-order effective Lagrangians of only two types of seagull 
term with adjustable coefficients, corresponding to the two polarisabilities. 
These Lagrangians do contain other seagull terms at
this order, but these all have coefficents that are fixed by the requirements 
of covariance. In particular, there is one that contributes to $T_1(0,Q^2)$ 
at order $Q^2$, but only through the ``non-pole" piece of the Born amplitude.
It is important keep this term in the Born amplitude, rather than 
subsuming it into the structure part \cite{cv11a}, as otherwise the magnetic 
polarisability in the LET Eq.~(\ref{eq:let1}) will not correspond to its 
conventional definition.

Our results leave no room for any large additional polarisability effect arising 
from off-shell form factors, as suggested by Miller et al.~\cite{mtcr11}.
The form factor appearing in $\overline T_1(0,Q^2)$ falls off with $Q^2$ and 
its mass parameter has a natural size for a pion-loop effect. We 
see no sign of any rapid growth at low $Q^2$ that could lead to a large 
contribution to the Lamb shift.

There is thus no evidence that effects of proton polarisability can explain 
the difference between the proton charge radius as recently determined from 
the muonic Lamb shift \cite{pohl10} and that obtained from other experiments 
\cite{codata08,a110}. The resolution of this puzzle must lie elsewhere, 
perhaps in re-analyses of the older experiments as suggested by Lorenz 
\textit{et al.}~\cite{lhm12}. 

\section*{Acknowledgments}

We are grateful to G. Miller for discussions that prompted this work. This work was 
supported by the UK STFC under grants ST/F012047/1 and ST/J000159/1.

\appendix*

\section{Full amplitude}

Our result for the the V$^2$CS amplitude as a full function of $Q^2$ can
be expressed in the form
\begin{equation}
\overline T_1=\overline T_1^{(3)}+\overline T_1^{(4)}+\overline T_1^{(\Delta)},
\end{equation}
where the third-order, fourth-order and $\Delta$-pole contributions are:
\begin{widetext}
\begin{eqnarray}
\overline T_1^{(3)}&=&\frac{e^2g_{\scriptscriptstyle A}^2m_\pi}{16 \pi  f_\pi^2}
\left[1-\frac{2 }{\sqrt{Q^2}}\tan^{-1}\left(\frac{\sqrt{Q^2}}{2}\right)\right]
=4\pi\,\frac{e^2g_{\scriptscriptstyle A}^2}{768 \pi^2m_\pi  f_\pi^2}\, m_\pi^2Q^2
+{\cal O}(Q^4)\,,\\
\noalign{\vspace{5pt}}
\overline T_1^{(4)}&=&4\pi\delta\beta\,  m_\pi^2Q^2\nonumber\\
\noalign{\vspace{5pt}}
&&+\,\frac{e^2m_\pi^2}{288 \pi ^2 f_\pi^2}
\Biggl\{12 (c_3-2c_1) (Q^2+12)+16 c_2 (Q^2+3)
+3\,\frac{g_{\scriptscriptstyle A}^2}{M_{\scriptscriptstyle\rm N}}\,
\Bigl(6-(12 \mu_s+13) Q^2\Bigr)\nonumber\\
\noalign{\vspace{5pt}}
&&\hspace{2cm}-\,\frac{12}{\sqrt{Q^2 (Q^2+4)}}
\tanh^{-1}\left(\sqrt{\frac{Q^2}{Q^2+4}}\right) 
\Biggl[12 (c_3-2c_1) (Q^2+4)+c_2 (Q^2+4)^2\nonumber\\
\noalign{\vspace{5pt}}
&&\hspace{8.5cm}-3\,\frac{g_{\scriptscriptstyle A}^2}{M_{\scriptscriptstyle\rm N}} 
\Bigl((\mu_s+1) Q^2(Q^2+4)-2\Bigr)\Biggr]\Biggr\},\\
\noalign{\vspace{5pt}}
\overline T_1^{(\Delta)}&=&\frac{2e^2}{(M_{\scriptscriptstyle \Delta}^2
-M_{\scriptscriptstyle\rm N}^2)
(M_{\scriptscriptstyle \Delta}+M_{\scriptscriptstyle\rm N})}\,
m_\pi^2 Q^2\left\{g_{\scriptscriptstyle M}^2
+g_{\scriptscriptstyle M}\,\frac{ g_{\scriptscriptstyle A} 
g_{\pi{\scriptscriptstyle\text{N}\Delta}}
m_\pi(M_{\scriptscriptstyle \Delta}+M_{\scriptscriptstyle\rm N}) 
\left[(Q^2+4) \tan^{-1}   \left(\frac{\sqrt{Q^2}}{2}\right)
-2 \sqrt{Q^2}\right]}{96   \pi f_\pi^2  \sqrt{Q^2}}\right\}.\nonumber\\
&&
\end{eqnarray}
\end{widetext}
where $Q^2$ has been expressed in units of $m_\pi^2$. Here $\delta \beta$ denotes
the sum of fourth-order loop and counterterm contributions to $\beta$. The 
subsequent terms in $\overline T_1^{(4)}$ are of order $Q^4$ and higher, and 
hence contribute to the form factor that is needed for the Lamb shift. 
The value of $\delta \beta$ is independent of the choice of renormalisation 
scale and it is adjusted so that the complete term proportional to $Q^2$ in 
$\overline T_1$ gives the observed magnetic polarisability, as 
determined by a fit to Compton scattering data in Ref.~\cite{gmpf12}.


\begin{thebibliography}{99}
\bibitem{pohl10}R. Pohl \textit{et al.}, Nature \textbf{466}, 213 (2010).
\bibitem{codata08}P. J. Mohr, B. N. Taylor and D. B. Newell, Rev. Mod. Phys.
\textbf{80}, 633 (2008) [arXiv:0801.0028].
\bibitem{a110}J. C. Bernauer \textit{et al.} (A1 Collaboration), Phys. Rev. Lett. 
\textbf{105}, 242001 (2010) [arXiv:1007.5076].
\bibitem{jen11a}U. D. Jentschura, Ann. Phys. \textbf{326}, 500 (2011)
[arXiv:1011.5275].
\bibitem{jen11b}U. D. Jentschura, Ann. Phys. \textbf{326}, 516 (2011) 
[arXiv:1011.5453].
\bibitem{bor12}E. Borie, Ann. Phys. \textbf{327}, 733 (2012) [arXiv:1103.1772]. 
\bibitem{codata12}P. J. Mohr, B. N. Taylor and D. B. Newell, arXiv:1203.5425.
\bibitem{bt76}J. Bernab\'eu and R. Tarrach, Ann. Phys. \textbf{102}, 323 (1976)
\bibitem{bj73}J. Bernab\'eu and C. Jarlskog, Nucl. Phys. B \textbf{60}, 347 (1973).
\bibitem{pach99}K. Pachucki, Phys. Rev. A \textbf{60}, 3593 (1999) 
[arXiv:physics/9906002].
\bibitem{skk96}S. Scherer, A. Yu. Korchin and J. H. Koch, Phys. rev. C \textbf{54}, 
904 (1996) [arXiv:nucl-th/9605030].
\bibitem{dkms97}D. Drechsel, G. Knoechlein, A. Metz and S. Scherer, Phys. Rev. 
C \textbf{55}, 424 (1997) [arXiv:nucl-th/9608061]. 
\bibitem{fs98}H. W. Fearing and S. Scherer, Few-Body Syst. \textbf{23}, 111 (1998)
[arXiv:nucl-th/9607056].
\bibitem{mart06}A. P. Martynenko, Phys. Atom. Nucl. \textbf{69}, 1309 (2006)
[arXiv:hep-ph/0509236]. 
\bibitem{zee72}A. Zee, Phys. Rept. \textbf{3}, 127 (1972)
\bibitem{gl82}J. Gasser and H. Leutwyler, Phys. Rept. \textbf{87}, 77 (1982).
\bibitem{wlcm12}A. Walker-Loud, C. E. Carlson and G. A. Miller, arXiv:1203.0254.
\bibitem{mtcr11}G. A. Miller, A. W. Thomas, J. D. Carroll and J. Rafelski,
Phys. Rev. A \textbf{84}, 020101 (2011) [arXiv:1101.4073].
\bibitem{cv11a}C. E. Carlson and M. Vanderhaeghen, Phys. Rev. A \textbf{84}, 
020102 (2011) [arXiv:1101.5965].
\bibitem{hp11}R. J. Hill and G. Paz, Phys. Rev. Lett. \textbf{107}, 160402 (2011) 
[arXiv:1103.4617].
\bibitem{cv11b}C. E. Carlson and M. Vanderhaeghen, arXiv:1109.3779.
\bibitem{np08}D. Nevado and A. Pineda, Phys. Rev. C \textbf{77}, 035202 (2008)
[arXiv:0712.1294].
\bibitem{bksm93}V. Bernard, N. Kaiser, A. Schmidt and U.-G. Meissner,
Phys. Lett. B \textbf{319}, 269 (1993) [arXiv:hep-ph/9309211];
V. Bernard, N. Kaiser, U.-G. Meissner and A. Schmidt,
Z. Phys. A \textbf{348}, 317 (1994) [arXiv:hep-ph/9311354].
\bibitem{bs92}M. N. Butler and M. J. Savage, Phys. Lett. \textbf{294}, 369 (1992) 
[arXiv:hep-ph/9209204].
\bibitem{hhk97}T. R. Hemmert, B. R. Holstein and J. Kambor, Phys. Rev.
D \textbf{55}, 5598 (1997) [arXiv:hep-ph/9612374].
\bibitem{pp03}V. Pascalutsa and D. R. Phillips, Phys. Rev. C \textbf{67}, 055202
(2003) [arXiv:nucl-th/0212024].
\bibitem{pin05}A. Pineda, Phys. Rev. C \textbf{71}, 065205 (2005) [arXiv:hep-ph/0412142].
\bibitem{bkm93}V. Bernard, N. Kaiser and U.-G. Meissner,
Phys. Rev. D \textbf{48}, 3062  (1993) [arXiv:hep-ph/9212257].
\bibitem{jko00}X. Ji, C.-W. Kao and J. Osborne, Phys. Lett. B \textbf{472}, 
1 (2000) [arXiv:hep-ph/9910256].
\bibitem{bhm02}V. Bernard, T. R. Hemmert and U.-G. Meissner,
Phys. Lett. B \textbf{545}, 105 (2002) [arXiv:hep-ph/0203167];  
Phys. Rev. D \textbf{67}, 076008 (2003) [arXiv:hep-ph/0212033].
\bibitem{gmpf12}H. W. Griesshammer, J. A. McGovern, D. R. Phillips and G. Feldman,
Prog. Part. Nucl. Phys. \textbf{67}, 841 (2012) [arXiv:1203.6834]. 
\bibitem{col79}J. C. Collins, Nucl. Phys. B \textbf{149}, 90 (1979).
\bibitem{bt68}W. A. Bardeen and W.-K. Tung, Phys. Rev. \textbf{173}, 1423 (1968). 
\bibitem{tar75}R. Tarrach, Nuovo Cim, \textbf{28A}, 409 (1975).
\bibitem{bkm95}V. Bernard, N. Kaiser and U.-G. Meissner, Int. J. Mod. Phys. E
\textbf{4}, 193 (1995).
\bibitem{lm92}M. E. Luke and A. V. Manohar, Phys. Lett. B \textbf{286}, 348 (1992)
[arXiv:hep-ph/9205228].
\bibitem{man97}A. V. Manohar, Phys. Rev. D \textbf{56}, 230 (1997)
[arXiv:hep-ph/9701294].
\bibitem{fmms00}N. Fettes, U.-G. Meissner, M. Mojzis and S. Steininger,
Ann. Phys. \textbf{283}, 273 (2000); erratum \textit{ibid.} \textbf{288}, 
249 (2001).
\bibitem{bkm92}V. Bernard, N. Kaiser and U.-G. Meissner, Nucl. Phys. B \textbf{373},
346 (1992).
\bibitem{mcg01}J. A. McGovern, Phys. Rev. C \textbf{63}, 064608 (2001); erratum
\textit{ibid.} \textbf{66}, 039902 (2002) [arXiv:nucl-th/0101057].
\bibitem{bjm01}M. C. Birse, X. Ji and J. A. McGovern Phys. Rev. Lett. 
\textbf{86}, 3204 (2001) [arXiv:nucl-th/0011054].
\bibitem{pv05}V.~Pascalutsa and M.~Vanderhaeghen,
Phys. Rev.  D {\bf 73}, 034003 (2006) [arXiv:hep-ph/0512244].
\bibitem{pvy06}V.~Pascalutsa, M.~Vanderhaeghen and S.~N.~Yang,
Phys. Rept. {\bf 437}, 125 (2007) [arXiv:hep-ph/0609004].
\bibitem{ber07}V.~Bernard, Prog. Part. Nucl. Phys. {\bf 60}, 82 (2008)
[arXiv:0706.0312].
\bibitem{rts03}M.~C.~M.~Rentmeester, R.~G.~E.~Timmermans and J.~J.~de Swart,
Phys. Rev. C {\bf 67}, 044001 (2003) [arXiv:nucl-th/0302080].
\bibitem{hghp04}R. P. Hildebrandt, H. W. Grießhammer, T. R. Hemmert and B. Pasquini,
Eur. Phys. J. A \textbf{20}, 293 (2004) [arXiv:nucl-th/0307070].
\bibitem{vz78}A. I. Vainstein and V. I. Zakharov, Phys. Lett. B \textbf{72},
368 (1978).
\bibitem{cg84}C. E. Carlson and F. Gross, Phys. Rev. Lett. \textbf{53}, 127
(1984).
\bibitem{bkkm92}V. Bernard, N. Kaiser, J. Kambor and U.-G. Meissner, 
Nucl. Phys. B \textbf{388}, 287 (1992).
\bibitem{bfhm98}V. Bernard, H. W. Fearing, T. R. Hemmert and U.-G. Meissner, 
Nucl. Phys. A \textbf{635}, 121 (1998) [arXiv:hep-ph/9801297].
\bibitem{pach96}K. Pachucki, Phys. Rev. A \textbf{53}, 2093 (1996).
\bibitem{amt07}J. Arrington, W. Melnitchouk and J. A. Tjon, Phys. Rev. C
\textbf{76}, 035205 (2007) [arXiv:0707.1861].
\bibitem{lhm12}I. T. Lorenz, H.-W. Hammer and U.-G. Meissner, arXiv:1205:6628.

\end{thebibliography}
\end{document}